\RequirePackage{lineno}
\RequirePackage{xspace}
\documentclass[aps,prl,twocolumn,showpacs,superscriptaddress,groupedaddress]{revtex4}  % for review and submission

\usepackage{graphicx}  % needed for figures
\usepackage{dcolumn}   % needed for some tables
\usepackage{bm}        % for math
\usepackage{amssymb}   % for math
\usepackage{slashed}

\newcommand{\Dzero}{D$0$\xspace}

\def\bc {\begin{center}}
\def\ec {\end{center}}
\def\beq {\begin{equation}}
\def\eeq {\end{equation}}

\def\NN  {\ensuremath{{{\sl N\kern -0.1em N}}\xspace}}

\def \tanb  {\ensuremath{\mathrm{\tan \beta}}\xspace}
\def \taum {\ensuremath{\mu}\xspace}
\def \tauh {\ensuremath{\tau_h}\xspace}
\def \b        {\ensuremath{b}\xspace}
\def \invfb       {\ensuremath{\mathrm{fb}^{-1}}}

\def\pt {\ensuremath{p_T}\xspace}
\def \pttau {\ensuremath{ p_T^{\tauh} }\xspace}
\def \ptmu {\ensuremath{ p_T^{\taum}}\xspace}

\def \vpttau {\ensuremath{  \vec{p}_T^{\ \tauh}    }\xspace}
\def \vptmu {\ensuremath{  \vec{p}_T^{\ \taum}   }\xspace}

\def\met    {\ensuremath{\slashed{E}_T}\xspace}
\def\vmet    {\ensuremath{\vec{\slashed{E}}_T}\xspace}

\def\mD {\ensuremath{\mathcal{D}}\xspace}

\def\Dqcd {\ensuremath{\mD_{\text{MJ}}}\xspace}
\def\Dtt {\ensuremath{\mD_{\t\tbar}}\xspace}
\def\Df {\ensuremath{\mD_{f}}\xspace}

\def\RMJ{\ensuremath{R_{\text{iso}/\overline{\text{iso}}}}\xspace}

\def\mhat{\ensuremath{M_{\text{hat}}}\xspace}

\def\BR         {{\ensuremath{\cal B}\xspace}}

\def\proton      {\ensuremath{p}\xspace}
\def\antiproton  {\ensuremath{\overline p}\xspace}
\def\c     {\ensuremath{c}\xspace}

\def\b     {\ensuremath{b}\xspace}
\def\bbar  {\ensuremath{\overline b}\xspace}

\def\t     {\ensuremath{t}\xspace}
\def\tbar  {\ensuremath{\overline t}\xspace}
\def\tbar  {\ensuremath{\overline t}\xspace}
\def\ttbar {\ensuremath{t\overline t}\xspace}

\def\pizs  {\ensuremath{\pi^0\mbox\,\rm{s}}\xspace}

\newcommand{\tev}{\ensuremath{\mathrm{\,Te\kern -0.1em V}}\xspace}
\newcommand{\gev}{\ensuremath{\mathrm{\,Ge\kern -0.1em V}}\xspace}
\newcommand{\mev}{\ensuremath{\mathrm{\,Me\kern -0.1em V}}\xspace}
\newcommand{\kev}{\ensuremath{\mathrm{\,ke\kern -0.1em V}}\xspace}
\newcommand{\ev}{\ensuremath{\mathrm{\,e\kern -0.1em V}}\xspace}
\newcommand{\gevc}{\ensuremath{{\mathrm{\,Ge\kern -0.1em V\!/}c}}\xspace}
\newcommand{\mevc}{\ensuremath{{\mathrm{\,Me\kern -0.1em V\!/}c}}\xspace}
\newcommand{\gevcc}{\ensuremath{{\mathrm{\,Ge\kern -0.1em V\!/}c^2}}\xspace}
\newcommand{\mevcc}{\ensuremath{{\mathrm{\,Me\kern -0.1em V\!/}c^2}}\xspace}

\def\cm   {\ensuremath{{\rm \,cm}}\xspace}

\newcommand{\jprlBase}       {Phys.\ Rev.\ Lett.\xspace}
\newcommand{\jprBase}        {Phys.\ Rev.\xspace}
\newcommand{\jplBase}        {Phys.\ Lett.\xspace}

\newcommand{\nimBaseD}       {Nucl.\ Instrum.\ Methods\ Phys.\ Res.\xspace}

\newcommand{\nima}      [1]  {\nimBaseD~A~{\bf #1}}

\newcommand{\plb}       [1]  {\jplBase\ B~{\bf #1}}

\newcommand{\jprl}      [1]  {\jprlBase\ {\bf #1}}
\newcommand{\jprd}      [1]  {\jprBase\ D~{\bf #1}}
\begin{document}

% the following line is for submission, including submission to the arXiv!!
\hspace{5.2in} \mbox{FERMILAB-PUB-11-293-E}

\title{Search for neutral Higgs bosons decaying to tau pairs produced in association with \boldmath{$b$} quarks in \boldmath{$\proton\antiproton$} collisions at \boldmath{$\sqrt{s}=1.96~\tev$}}
\affiliation{Universidad de Buenos Aires, Buenos Aires, Argentina}
\affiliation{LAFEX, Centro Brasileiro de Pesquisas F{\'\i}sicas, Rio de Janeiro, Brazil}
\affiliation{Universidade do Estado do Rio de Janeiro, Rio de Janeiro, Brazil}
\affiliation{Universidade Federal do ABC, Santo Andr\'e, Brazil}
\affiliation{Instituto de F\'{\i}sica Te\'orica, Universidade Estadual Paulista, S\~ao Paulo, Brazil}
\affiliation{Simon Fraser University, Vancouver, British Columbia, and York University, Toronto, Ontario, Canada}
\affiliation{University of Science and Technology of China, Hefei, People's Republic of China}
\affiliation{Universidad de los Andes, Bogot\'{a}, Colombia}
\affiliation{Charles University, Faculty of Mathematics and Physics, Center for Particle Physics, Prague, Czech Republic}
\affiliation{Czech Technical University in Prague, Prague, Czech Republic}
\affiliation{Center for Particle Physics, Institute of Physics, Academy of Sciences of the Czech Republic, Prague, Czech Republic}
\affiliation{Universidad San Francisco de Quito, Quito, Ecuador}
\affiliation{LPC, Universit\'e Blaise Pascal, CNRS/IN2P3, Clermont, France}
\affiliation{LPSC, Universit\'e Joseph Fourier Grenoble 1, CNRS/IN2P3, Institut National Polytechnique de Grenoble, Grenoble, France}
\affiliation{CPPM, Aix-Marseille Universit\'e, CNRS/IN2P3, Marseille, France}
\affiliation{LAL, Universit\'e Paris-Sud, CNRS/IN2P3, Orsay, France}
\affiliation{LPNHE, Universit\'es Paris VI and VII, CNRS/IN2P3, Paris, France}
\affiliation{CEA, Irfu, SPP, Saclay, France}
\affiliation{IPHC, Universit\'e de Strasbourg, CNRS/IN2P3, Strasbourg, France}
\affiliation{IPNL, Universit\'e Lyon 1, CNRS/IN2P3, Villeurbanne, France and Universit\'e de Lyon, Lyon, France}
\affiliation{III. Physikalisches Institut A, RWTH Aachen University, Aachen, Germany}
\affiliation{Physikalisches Institut, Universit{\"a}t Freiburg, Freiburg, Germany}
\affiliation{II. Physikalisches Institut, Georg-August-Universit{\"a}t G\"ottingen, G\"ottingen, Germany}
\affiliation{Institut f{\"u}r Physik, Universit{\"a}t Mainz, Mainz, Germany}
\affiliation{Ludwig-Maximilians-Universit{\"a}t M{\"u}nchen, M{\"u}nchen, Germany}
\affiliation{Fachbereich Physik, Bergische Universit{\"a}t Wuppertal, Wuppertal, Germany}
\affiliation{Panjab University, Chandigarh, India}
\affiliation{Delhi University, Delhi, India}
\affiliation{Tata Institute of Fundamental Research, Mumbai, India}
\affiliation{University College Dublin, Dublin, Ireland}
\affiliation{Korea Detector Laboratory, Korea University, Seoul, Korea}
\affiliation{CINVESTAV, Mexico City, Mexico}
\affiliation{Nikhef, Science Park, Amsterdam, the Netherlands}
\affiliation{Radboud University Nijmegen, Nijmegen, the Netherlands and Nikhef, Science Park, Amsterdam, the Netherlands}
\affiliation{Joint Institute for Nuclear Research, Dubna, Russia}
\affiliation{Institute for Theoretical and Experimental Physics, Moscow, Russia}
\affiliation{Moscow State University, Moscow, Russia}
\affiliation{Institute for High Energy Physics, Protvino, Russia}
\affiliation{Petersburg Nuclear Physics Institute, St. Petersburg, Russia}
\affiliation{Instituci\'{o} Catalana de Recerca i Estudis Avan\c{c}ats (ICREA) and Institut de F\'{i}sica d'Altes Energies (IFAE), Barcelona, Spain}
\affiliation{Stockholm University, Stockholm and Uppsala University, Uppsala, Sweden}
\affiliation{Lancaster University, Lancaster LA1 4YB, United Kingdom}
\affiliation{Imperial College London, London SW7 2AZ, United Kingdom}
\affiliation{The University of Manchester, Manchester M13 9PL, United Kingdom}
\affiliation{University of Arizona, Tucson, Arizona 85721, USA}
\affiliation{University of California Riverside, Riverside, California 92521, USA}
\affiliation{Florida State University, Tallahassee, Florida 32306, USA}
\affiliation{Fermi National Accelerator Laboratory, Batavia, Illinois 60510, USA}
\affiliation{University of Illinois at Chicago, Chicago, Illinois 60607, USA}
\affiliation{Northern Illinois University, DeKalb, Illinois 60115, USA}
\affiliation{Northwestern University, Evanston, Illinois 60208, USA}
\affiliation{Indiana University, Bloomington, Indiana 47405, USA}
\affiliation{Purdue University Calumet, Hammond, Indiana 46323, USA}
\affiliation{University of Notre Dame, Notre Dame, Indiana 46556, USA}
\affiliation{Iowa State University, Ames, Iowa 50011, USA}
\affiliation{University of Kansas, Lawrence, Kansas 66045, USA}
\affiliation{Kansas State University, Manhattan, Kansas 66506, USA}
\affiliation{Louisiana Tech University, Ruston, Louisiana 71272, USA}
\affiliation{Boston University, Boston, Massachusetts 02215, USA}
\affiliation{Northeastern University, Boston, Massachusetts 02115, USA}
\affiliation{University of Michigan, Ann Arbor, Michigan 48109, USA}
\affiliation{Michigan State University, East Lansing, Michigan 48824, USA}
\affiliation{University of Mississippi, University, Mississippi 38677, USA}
\affiliation{University of Nebraska, Lincoln, Nebraska 68588, USA}
\affiliation{Rutgers University, Piscataway, New Jersey 08855, USA}
\affiliation{Princeton University, Princeton, New Jersey 08544, USA}
\affiliation{State University of New York, Buffalo, New York 14260, USA}
\affiliation{Columbia University, New York, New York 10027, USA}
\affiliation{University of Rochester, Rochester, New York 14627, USA}
\affiliation{State University of New York, Stony Brook, New York 11794, USA}
\affiliation{Brookhaven National Laboratory, Upton, New York 11973, USA}
\affiliation{Langston University, Langston, Oklahoma 73050, USA}
\affiliation{University of Oklahoma, Norman, Oklahoma 73019, USA}
\affiliation{Oklahoma State University, Stillwater, Oklahoma 74078, USA}
\affiliation{Brown University, Providence, Rhode Island 02912, USA}
\affiliation{University of Texas, Arlington, Texas 76019, USA}
\affiliation{Southern Methodist University, Dallas, Texas 75275, USA}
\affiliation{Rice University, Houston, Texas 77005, USA}
\affiliation{University of Virginia, Charlottesville, Virginia 22901, USA}
\affiliation{University of Washington, Seattle, Washington 98195, USA}
\author{V.M.~Abazov} \affiliation{Joint Institute for Nuclear Research, Dubna, Russia}
\author{B.~Abbott} \affiliation{University of Oklahoma, Norman, Oklahoma 73019, USA}
\author{B.S.~Acharya} \affiliation{Tata Institute of Fundamental Research, Mumbai, India}
\author{M.~Adams} \affiliation{University of Illinois at Chicago, Chicago, Illinois 60607, USA}
\author{T.~Adams} \affiliation{Florida State University, Tallahassee, Florida 32306, USA}
\author{G.D.~Alexeev} \affiliation{Joint Institute for Nuclear Research, Dubna, Russia}
\author{G.~Alkhazov} \affiliation{Petersburg Nuclear Physics Institute, St. Petersburg, Russia}
\author{A.~Alton$^{a}$} \affiliation{University of Michigan, Ann Arbor, Michigan 48109, USA}
\author{G.~Alverson} \affiliation{Northeastern University, Boston, Massachusetts 02115, USA}
\author{G.A.~Alves} \affiliation{LAFEX, Centro Brasileiro de Pesquisas F{\'\i}sicas, Rio de Janeiro, Brazil}
% ...
%
% visitor_addresses.tex                        2 June 2011
%  available symbols are:
%  $\ast, \dag, \ddag, \S, \P, $\|$, $\ast\ast$, \dag\dag, \ddag\ddag ,\#
%
\collaboration{The D0 Collaboration\footnote{with visitors from
%{alton}
$^{a}$Augustana College, Sioux Falls, SD, USA,
%{burdin}
$^{b}$The University of Liverpool, Liverpool, UK,
%{haas,partridge}
$^{c}$SLAC, Menlo Park, CA, USA,
%{hesketh}
$^{d}$University College London, London, UK,
%{luna-garcia}
$^{e}$Centro de Investigacion en Computacion - IPN, Mexico City, Mexico,
%{podesta-lerma}
$^{f}$ECFM, Universidad Autonoma de Sinaloa, Culiac\'an, Mexico,
and 
%{weber}
$^{g}$Universit{\"a}t Bern, Bern, Switzerland.
%{garcia-guerra}
%$^{?}$UPIITA-IPN, Mexico City, Mexico,
%{hooper}
%$^{?}$Visitor from Bradley University, Peoria, IL, USA.
%{kozminski}
%$^{?}$}Visitor from Lewis University, Romeoville, IL, USA.
%{deceased}
$^{\ddag}$Deceased.
}} \noaffiliation
\vskip 0.25cm

\date{June 24, 2011}

\begin{abstract}
We report results from a search for neutral Higgs bosons produced in association with  $b$ quarks using data recorded by the \Dzero experiment at the Fermilab Tevatron Collider and corresponding to an integrated luminosity of $7.3~\invfb$.  This production mode can be enhanced in several extensions of the standard model (SM) such as in its minimal supersymmetric extension (MSSM) at high \tanb.  We search for Higgs bosons decaying to tau pairs with one tau decaying to a muon and neutrinos and the other to hadrons.
The data are found to be consistent with SM expectations, and we set upper limits on the cross section times branching ratio in the Higgs boson mass range from $90$ to $320~\gevcc$.  We interpret our result in the MSSM parameter space, excluding \tanb values down to 25 for Higgs boson masses below $170~\gevcc$.
\end{abstract}

\maketitle

In contrast to the standard model (SM), where only one Higgs boson doublet breaks the $SU$(2) symmetry,  there are two Higgs boson doublets in the minimal supersymmetric standard model (MSSM)~\cite{mssm}. This leads to five physical Higgs bosons remaining after electroweak symmetry breaking; three neutrals: $h$, $H$, and $A$, collectively denoted as $\phi$, and two charged, $H^\pm$. 
At the tree level, the mass spectrum of the Higgs bosons is determined by two parameters conventionally chosen to be  \tanb, the ratio of the two Higgs doublet vacuum expectation values, and $M_A$, the mass of the pseudoscalar Higgs boson $A$.  Although \tanb is a free parameter in the MSSM,  large values ($\tanb\gtrsim20$) 
are preferred. The top quark to bottom quark mass ratio suggests $\tanb\approx35$~\cite{cite:topbottom}, and the observed density of dark matter also points towards high \tanb values~\cite{cite:darkmatter}. At high values of \tanb, two of the neutral Higgs bosons ($A$ and $h$ or $H$) are approximately degenerate in mass. They share similar couplings to quarks, enhanced  by $\tanb$ compared to the SM couplings for down-type fermions, while the couplings to up-type fermions are suppressed. The enhancement of couplings to down-type fermions has several consequences. First,  the main decay modes of this Higgs boson pair are $\phi\to\b\bbar$ and $\phi\to\tau\tau$ with  branching ratios $\BR(\phi\to\b\bbar)\approx90\%$ and $\BR(\phi\to\tau\tau)\approx10\%$, respectively.  Their production in association with $b$ quarks is enhanced by approximately $\tan^2\beta$ compared to the SM, which could make this production rate measurable at a hadron collider.

Experiments at the CERN $e^+e^-$ Collider (LEP) excluded MSSM Higgs boson masses below $93~\gevcc$~\cite{cite:LEP_exclu}. The CDF and \Dzero collaborations at the Tevatron extended the exclusion to higher masses for high $\tanb$~\cite{cite:CDF_tautau2,cite:D0_tautau2,cite:D0_tautau3,cite:D0_bbb3,cite:D0_btautau2}.
More recently, similar searches were performed at the LHC~\cite{cite:CMS_tautau}. In this letter, we present a search for the process $\proton\antiproton\to\phi\b\to\tau\tau\b$ where one $\tau$ lepton (denoted $\tau_\mu$) decays via  $\tau\to\mu\nu_\mu\nu_\tau$ and the other (denoted $\tauh$) decays hadronically. 
This mode is complementary to the inclusive $\phi\to\tau\tau$~\cite{cite:CDF_tautau2,cite:D0_tautau2} and the $\phi\b\to\b\b\b$~\cite{cite:D0_bbb3} searches.
This is because in the former, the presence of $b$ quark(s) in the final state significantly  decreases the $Z$ boson background, while the latter has a larger branching ratio but suffers from a large multijet background and is more sensitive to the MSSM parameters.
This result is built on, and supersedes, our previous result based on $2.7~\invfb$ of integrated luminosity~\cite{cite:D0_btautau2}. In addition to the increase in luminosity, the sensitivity is improved by a refined treatment of systematic uncertainties,  higher-performance signal to background discriminants and a higher trigger efficiency. \\

%%%% dataset / detector / object reconstruction
The data considered in this analysis were recorded by the \Dzero detector, described in~\cite{run2det}, and correspond to an integrated luminosity of $7.3~\invfb$~\cite{d0lumi}. Events were recorded using a mixture of single high-\pt muon, jet, tau, muon plus jet, and muon plus tau triggers.  A %pure 
data sample of $Z\to\tau_\mu\tauh$ is employed to measure the efficiency of this inclusive trigger approach with respect to single muon triggers. This has been validated in $Z(\to\tau_\mu\tauh)$+jets events.  The overall trigger efficiency ranges between $80\%$ and $95\%$, depending on the kinematics and on the decay topology of the hadronically decaying $\tau$. We rely on all components of the \Dzero detector: tracking, calorimetry, and the muon system.  
%%%% description of the different objects reconstruction
Muons are identified from track segments reconstructed in the muon system that are spatially matched to reconstructed tracks in the inner tracking system, and muon system scintillator hits must be in time with the beam crossing to veto cosmic muons. 
Hadronic $\tau$ decays are reconstructed from energy deposits in the calorimeter~\cite{nntau} using a jet cone algorithm with radius $=0.3$~\cite{cone}. They are required to have associated tracks. The $\tau$ candidates are then split in three different categories which roughly correspond to one-prong $\tau$ decay with no \pizs ($\tauh$ type 1), one-prong decay with \pizs ($\tauh$ type 2), and multiprong decay ($\tauh$ type 3). In addition, we use a neural-network-based $\tauh$ identification ($\NN_{\tau}$) to separate quark and gluon jets from genuine hadronic $\tau$ decays~\cite{nntau}.  The $\NN_{\tau}$ is based on shower shape variables, isolation variables, and correlation variables between the tracking and the calorimeter energy measurements.  We require $\NN_\tau>0.9$ (0.95 for $\tauh$ type 3) which has an efficiency around $65\%$ while rejecting $\approx99\%$ of jets. 
Jets are identified as clusters of energy in the calorimeter reconstructed with the midpoint cone algorithm~\cite{cone} with radius $= 0.5$. Jet reconstruction and energy calibration are described in~\cite{cite:jetx}. All jets are required to pass a set of quality criteria and to have at least two reconstructed tracks originating from the $\proton\antiproton$ vertex matched within $\Delta R($track, jet-axis$)=\sqrt{(\Delta\eta)^2 +(\Delta\varphi)^2}<0.5$ (where $\eta$ is the pseudorapidity~\cite{pseudorapidity} and $\varphi$ the azimuthal angle). 
A neural network $b$-tagging algorithm~\cite{nnbtag} ($\NN_b$), with lifetime-based information involving the track impact parameters and secondary vertices as inputs, is used to identify jets from $b$ quarks. 
The missing transverse energy, \met, used to infer the presence of neutrinos, is reconstructed as the negative of the vector sum of the transverse energy of calorimeter cells with $|\eta|<3.2$. It is corrected for the energy scales of all reconstructed objects.

\begin{figure*}[t]
	\includegraphics[width=0.32\linewidth]{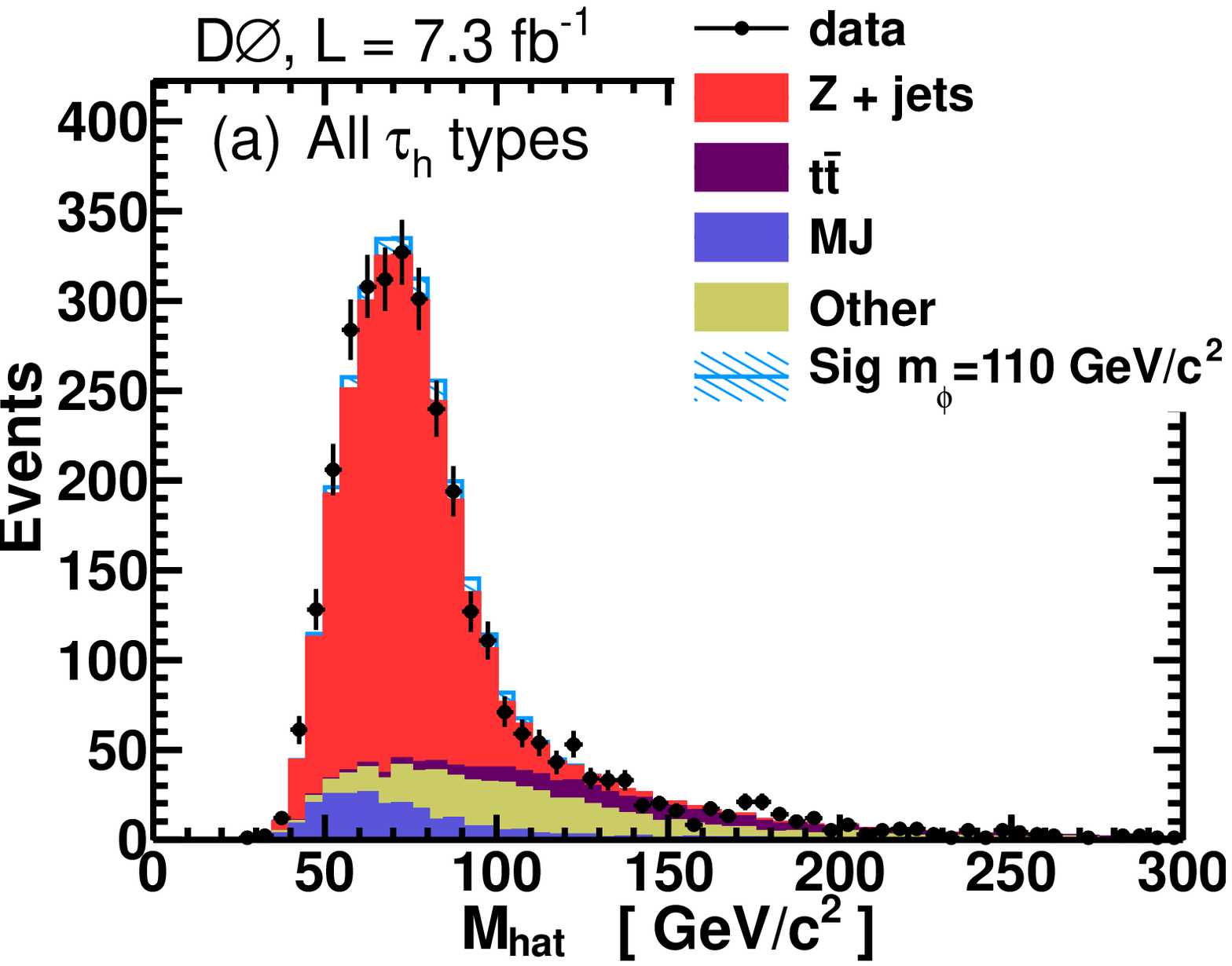}	
	\includegraphics[width=0.32\linewidth]{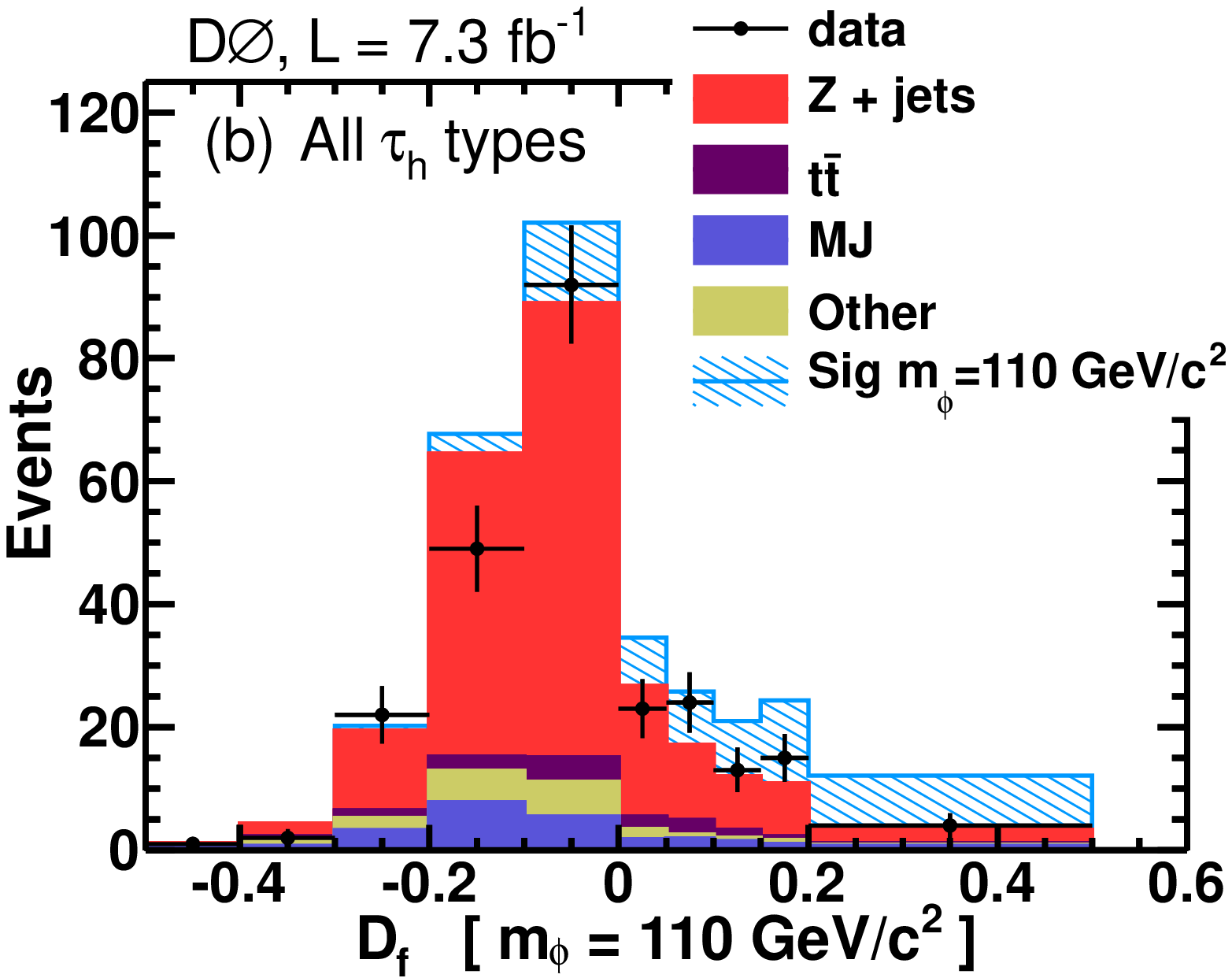}	
	\includegraphics[width=0.32\linewidth]{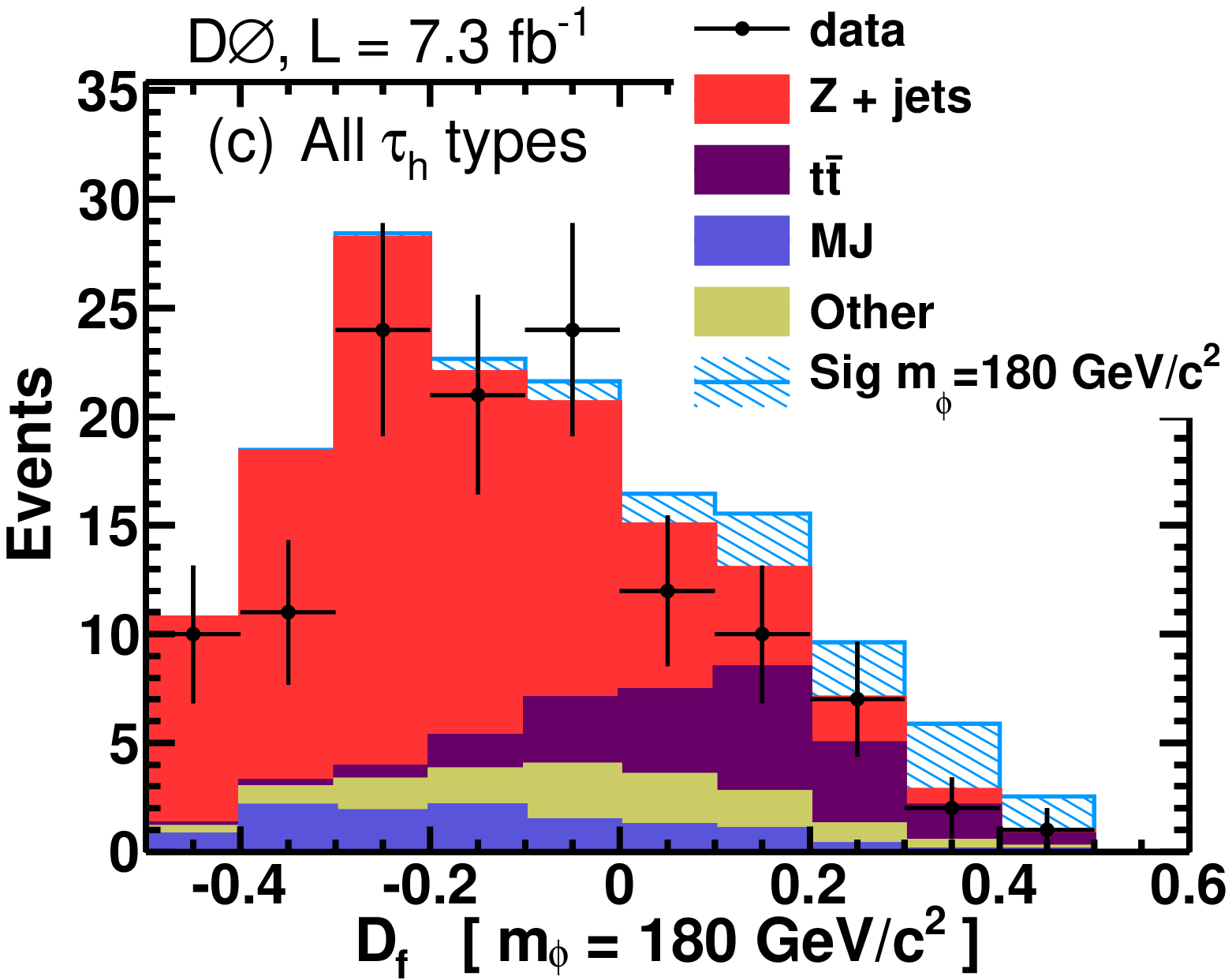}	
	\caption{(a) $\mhat$ distribution in the Pretag selection. (b) $\Df$ for a Higgs boson mass of $110~\gevcc$. (c) $\Df$ for a Higgs boson mass of $180~\gevcc$. The predicted signal is shown assuming the MSSM scenario described in the caption of Table~\ref{tab:nevts}.}
	\label{fig1}
\end{figure*}
%%%%% MC simulation
The leading order (LO) event generator {\sc pythia}~\cite{pythia} is used to generate $\phi\b$ production in the 5-flavor scheme, $g\b\to\phi\b$. To correct the cross section and the event kinematics to next-to-leading order (NLO), we use {\sc mcfm}~\cite{mcfm} to compute correction weights as a function of the leading $b$ quark \pt and $\eta$ in the range $p_T^b>12~\gevc$ and $|\eta^b|<5$.  The dominant backgrounds to this search are the production of $Z$+jets,  \ttbar and multijets (MJ), the latter being estimated from data. We also consider  $W$+jets and diboson ($WW$, $WZ$ and $ZZ$) production. Diboson events are simulated with {\sc pythia} while $Z$+jets, $W$+jets, and \ttbar samples are generated using {\sc alpgen}~\cite{alpgen} with {\sc pythia} for showering and hadronization. {\sc tauola}~\cite{tauola} is used for the decay of $\tau$ leptons;  $b$ hadron decays are modeled with {\sc evtgen}~\cite{evtgen}. The generated  samples are processed through a detailed simulation of the \Dzero detector based on {\sc geant}~\cite{geant}. The output is then combined with data events recorded during random beam crossings to model the effects of detector noise and pile-up energy from multiple interactions and different beam crossings. Finally, the same reconstruction algorithms as for data are used on the simulated events.
Corrections to the simulation are derived from data control samples and applied to object identification efficiencies, energy scales and resolutions, trigger efficiencies, and the longitudinal $\proton\antiproton$ vertex distribution. Signal, \ttbar, and diboson yields are determined from the product of the acceptance and detector efficiency (both determined from the simulation) multiplied by theoretical cross section times luminosity. For the dominant $Z\to\tau\tau$ background, the simulation is corrected by comparing a large sample of $Z\to\mu\mu$ events in data and in the simulation. This correction, measured in each jet multiplicity bin as a function of the $\phi^*$ event variable~\cite{phistar}, leading jet $\eta$, and leading $b$-tagged jet $\NN_b$, affects both the normalization and the kinematic distributions. For the $W$+jets background, the muon predominantly arises from the $W$ boson decay while the hadronic $\tau$ candidate is faked by a jet. While this background is estimated from the simulation, it is normalised to data using a $W(\to\mu\nu)+$jets control sample. \\
%%%% pretag sample
We define  a background-dominated sample, named Pretag in the following, to ensure our background modeling is correct. We select events with one reconstructed $\proton\antiproton$ vertex with at least three tracks, exactly one isolated muon (\taum), exactly one reconstructed hadronic tau (\tauh), and at least one jet. The muon is required to have a transverse momentum $\ptmu > 15~\gevc$, $|\eta^{\taum}| <1.6$, and to be isolated in the calorimeter and in the central tracking system, $\Delta R(\taum,\text{jet})>0.5$ relative to any reconstructed jet. The \tauh candidate  must satisfy $\pttau>10~\gevc$, $|\eta_{\tauh}| < 2.0$, $\Delta R(\tauh,\taum)>0.5$ relative to any muon, and \tauh tracks must not be shared with any reconstructed muons in the event. We also require the distance along the beam axis between \tauh and \taum  $\Delta z(\tauh,\taum) < 2~\cm$. Selected jets have $p_T^{\text{jet}}>15~\gevc$, $|\eta^{\text{jet}}|<2.5$, $\Delta R(\text{jet},\tauh)>0.5$. In addition, we require \tauh and \taum to have an opposite electric charge (OS) and a transverse mass $M_T(\taum,\met)< 60~\gevcc$ (100~\gevcc for $\tauh$ type 2). The transverse mass of $N$ reconstructed objects is defined as:
$$M_T(O_1,..,O_N) = \sqrt{ \sum_{O_i,O_j} p_T^{O_i} \cdot p_T^{O_j} \cdot \left[1- \cos \Delta \varphi(O_i,O_j) \right] },$$
where $\Delta \varphi(O_i,O_j)$ is the azimuthal angle between objects $O_i$ and $O_j$. Most of the MJ background is removed by the requirement $\Dqcd>0.1$ (0.2 for $\tauh$ type 3) where \Dqcd is a multivariate discriminant described below.
\begin{table}[t]
	\caption{Expected background yield, observed data yield,
	 and expected signal yields for the two selections described in the text with systematic uncertainties. 
	 The signal yields are given for the $m_h^{\text{max}}$ scenario ($\mu=+200\gev$ and $\tan\beta=40$). } 
	\label{tab:nevts}
	\begin{ruledtabular}
	\begin{tabular}{l  @{\hspace{0.5cm}} r@{$\ \pm$}r @{\hspace{0.5cm}}  r@{$\ \pm$}r}
 	& \multicolumn{2}{c}{Pretag} & \multicolumn{2}{c}{$b$-tagged} \\
	\hline
	$Z$+jets                                   & 2237.7 & 123.5 & 217.5 & 16.8\\	
	$t \bar{t}$   &   225.6 &  38.7 & 182.6 & 32.2\\
	MJ              &  225.0 &  39.6  & 28.4   & 4.8 \\
	Other           & 451.8  & 18.6 &  47.6  & 3.0 \\
	\hline
	Total background            &  3139.9 & 154.0  & 476.0 & 40.2 \\
	Data                 &  \multicolumn{2}{c}{ 3236 } &  \multicolumn{2}{c}{488}\\
Signal $m_\phi=110~\gevcc$ & \multicolumn{2}{c}{107.4} & \multicolumn{2}{c}{ 67.8 }\\
Signal $m_\phi=180~\gevcc$ & \multicolumn{2}{c}{24.0} & \multicolumn{2}{c}{ 15.0 }\\
	\end{tabular}
	\end{ruledtabular}
\end{table}
Finally, to improve the signal to background ratio, we  select a more restrictive $b$-tagged sample by demanding at least one jet to have $\NN_b>0.25$. This $b$-tag requirement  has an efficiency of $65\%$ for a probability of misidentifying a light parton jet as a $b$ jet of $5\%$. Table~\ref{tab:nevts} shows the predicted backgrounds, observed data yields, and expected signal yields in the pretag and $b$-tagged samples.

\begin{figure*}[t]
	\includegraphics[width=0.32\linewidth]{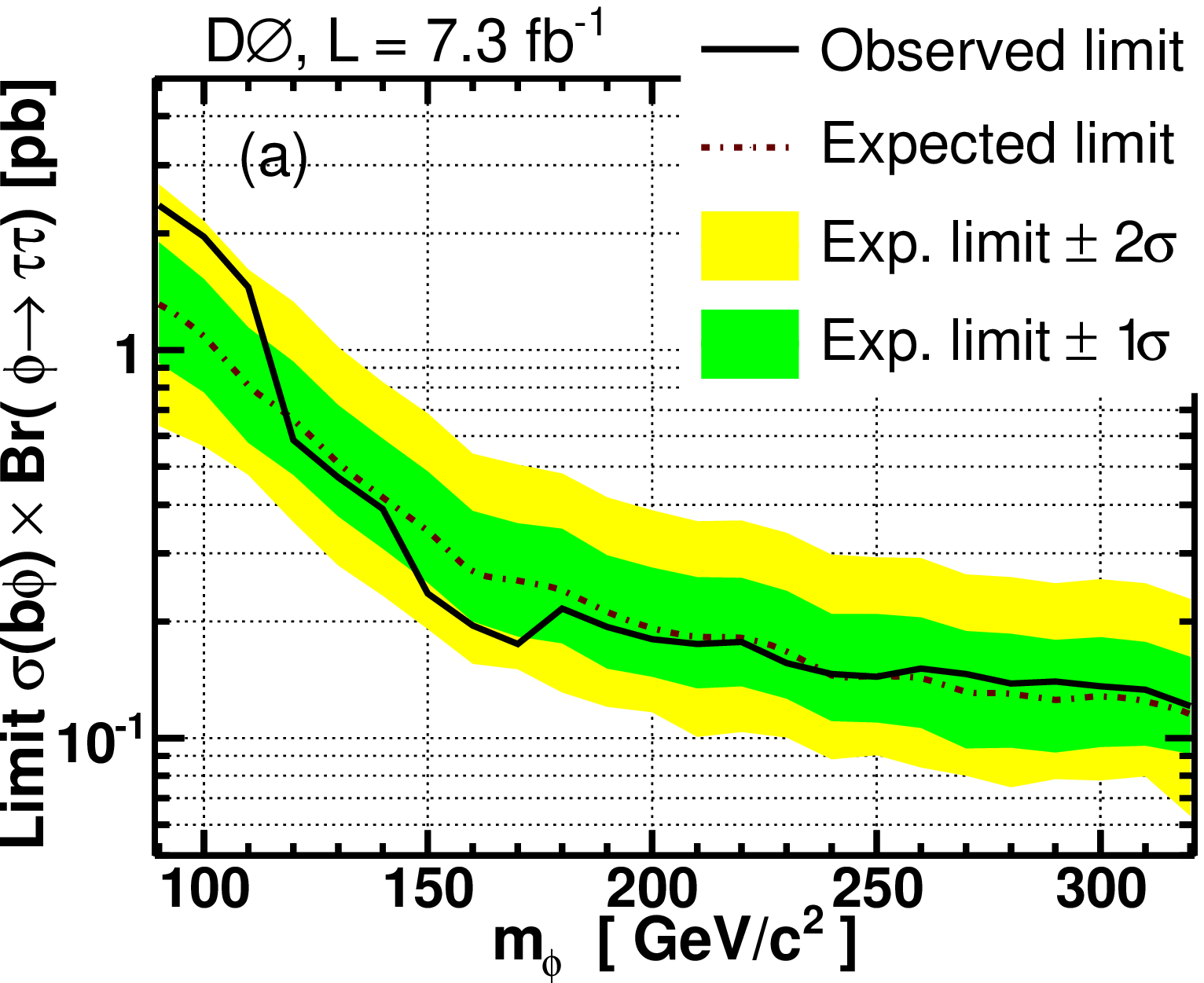}	
	\includegraphics[width=0.32\linewidth]{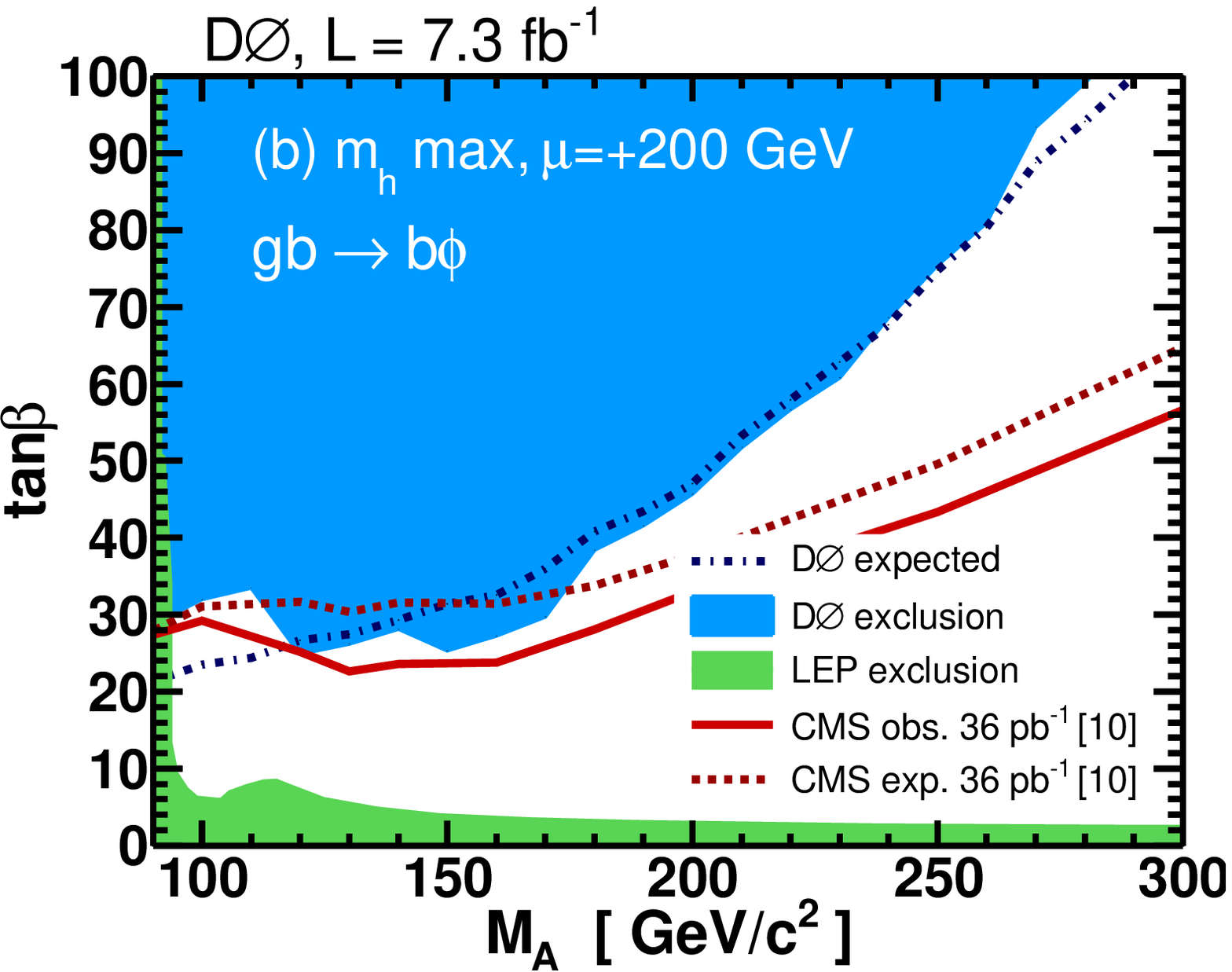}	
	\includegraphics[width=0.32\linewidth]{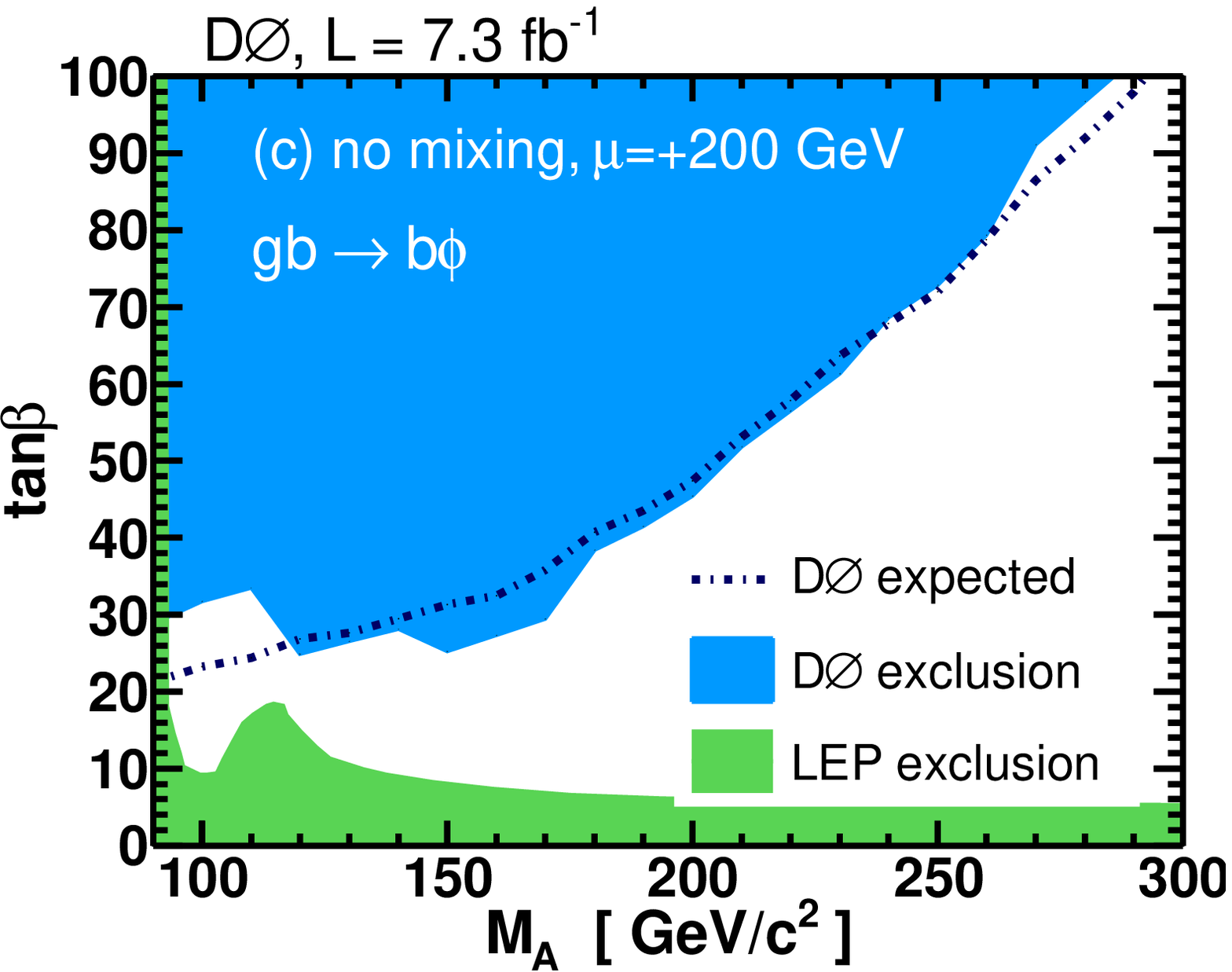}	
	\caption{(a) Model independent cross section times branching ratio limit as a function of $m_\phi$, (b) \tanb  vs $M_A$ limit in the MSSM $m_h^{\text{max}}$ scenario, and (c) in the MSSM no-mixing scenario.}
	\label{fig2}
\end{figure*}

The MJ background is estimated from control data samples. We define a MJ-enriched control sample with identical requirements as in the pretag and $b$-tagged signal samples but reversing the muon isolation criteria.  In a dedicated MJ sample obtained by requiring $\taum$ and $\tauh$ to have the same electric charge (SS), we measure the ratio of the probability for a MJ-event muon to appear isolated to the probability for a MJ-event muon to be non-isolated:   
$\RMJ\equiv  \mathcal{P}(\mu_{\mathrm{iso}} | \text{MJ}) / \mathcal{P}(\mu_{\overline{\mathrm{iso}}} | \text{MJ})$. 
The dependence on $\eta^{\tauh}$, $\pttau$, and leading-jet $\pt$ of \RMJ is taken into account. This \RMJ is then applied to events in the non-isolated-muon sample to predict the MJ background in the signal samples. 
%%% alternate MJ method
An alternate method is used to estimate the systematic uncertainty. For MJ events, we expect the correlation between the charge of \taum and \tauh to be small.  Therefore, we use a data sample that has the same selection as the $b$-tagged sample except that \taum and \tauh are SS. 
We subtract from this MJ-dominated SS sample the residual contribution from other SM backgrounds.
The number of MJ events in the OS signal sample is obtained by multiplying the SS sample yield by the OS:SS ratio, $1.07\pm0.01$, determined in the non-isolated-muon sample. The difference in normalization between the two methods is taken as a systematic uncertainty on the MJ contribution.

To further improve the signal to background discrimination, we use multivariate techniques. A first neural network \Dqcd is used to separate MJ background from the signal. Two \Dqcd discriminants are trained, one for $\tauh$ types 1 and 3, and another for $\tauh$ type~2. They are based on \ptmu, \pttau, \met, $|\Delta \varphi(\taum,\tauh)|$, $H_T\equiv \sum_{\text{jets}} p_T^{\text{jet}}$, $M_T(\text{All}O)$ (where the sum is performed over all objects), $\mhat$, and $M_{\text{col}}$. The quantity $\mhat$ is defined as 
$$\mhat\equiv\sqrt{\left( E^{\taum\tauh} - p_{z}^{\taum\tauh}  + \met  \right)^2 - | \vpttau + \vptmu + \vmet |^2},$$ 
where $E^{\taum\tauh}$ is the energy of the $\tauh\taum$ system, and  $p_z^{\taum\tauh}$ is its momentum along the beam axis. 
It represents the minimal center-of-mass energy consistent with a di-tau resonance decay. %%% can be removed
The quantity $M_{\text{col}}$ is the $\taum\tauh$ invariant mass assuming neutrinos are emitted along the $\tau$ decay axis~\cite{mcol}.  
To address the $\t\tbar$ background, we train  a neural network $\Dtt$ to discriminate against signals built from samples simulated at three consecutive Higgs boson masses, in order to increase the signal statistics.
It is constructed from the variables $|\Delta \varphi(\taum,\tauh)|$, $|\Delta\varphi(\taum,\met)|$, $H_T$, $H_T+\pttau+\ptmu$, \met, $M_T(\text{All}O)$, $M_T(\taum,\met)$, $\mhat$, $M_{\text{col}}$, $\mathcal{A}_T\equiv (\ptmu-\pttau)/\pttau$, and $N_{\text{jets}}$, the total number of jets in the event. Finally, for events satisfying $\Dtt>0.1$, we form a likelihood discriminant \Df which uses as input $\Dqcd$, $\Dtt$, $\NN_b$, and $\mhat$. 

%%% systematic uncertainties
Systematic uncertainties are divided in two categories: those affecting only the normalizations and those also  affecting the shapes of $\Df$ distributions. 
Those affecting the dominant $Z$+jets background modeling are evaluated with %dedicated 
$Z\to\mu\mu$ samples: $Z$+jets ($3.2\%$) and $Z$+\b-tagged jets  ($5\%$) normalizations, inclusive trigger efficiency ($3\%$) which also affects all other simulated processes,
%SM backgrounds and signals, 
$Z$ boson kinematics ($1\%$) which is shape-dependent. 
For non-$Z$ boson and non-MJ backgrounds, we consider the uncertainties affecting the normalization: luminosity ($6.1\%$), muon reconstruction efficiency ($2.9\%$), $\tauh$ reconstruction efficiency [($4$--$10)\%$],  single muon triggers efficiency ($1.3\%$), $\ttbar$ and diboson cross sections ($11\%$ and $7\%$), and the uncertainties affecting the shape of $\Df$: jet energy calibration ($\sim10\%$) and $b$-tagging ($\sim4\%$). The \tauh energy scale, and jet identification efficiencies have a negligible effect. The MJ background systematic uncertainties range from $10\%$ to $40\%$.

The predicted background, signal, and data distributions of $\mhat$ and $\Df$ discriminant are shown in Fig.~\ref{fig1}.  The \Df distributions are used as input to a significance calculation using the modified frequentist approach~\cite{cls,collie}. We do not observe any significant excess over the expected background. We first set model independent limits (assuming the Higgs boson width is negligible compared to the experimental resolution) at the $95\%$ C.L. on the signal cross section times branching fraction as a function of the Higgs boson mass; these are shown in Fig.~\ref{fig2}(a). These limits are then translated into the \tanb, $M_A$ plane for two MSSM benchmark scenarios~\cite{mssm_s}: the $m_h^{\text{max}}$ and no-mixing scenarios. The MSSM to SM signal ratio as well as the Higgs boson width are calculated with the {\sc feynhiggs} program~\cite{feynhiggs}. In this  interpretation, we further include systematic uncertainties on the signal production cross section ($15\%$)~\cite{cite:D0_bbb3}. We also take into account the Higgs boson width using the method described in~\cite{cite:D0_bbb3}. Figures~\ref{fig2}(b) and (c) present the limits for the two scenarios with the higgsino mass parameter $\mu=+200~\gev$. Numerical results and limits in other MSSM scenario are presented in~\cite{suppl}. We exclude a substantial region of the MSSM parameter space, especially at low $M_A$, and set the most stringent limit to date at a hadron collider, when using this final state.

% acknowledgement.tex                             6 April 2011
%
We thank the staffs at Fermilab and collaborating institutions,
and acknowledge support from the
DOE and NSF (USA);
CEA and CNRS/IN2P3 (France);
FASI, Rosatom and RFBR (Russia);
CNPq, FAPERJ, FAPESP and FUNDUNESP (Brazil);
DAE and DST (India);
Colciencias (Colombia);
CONACyT (Mexico);
KRF and KOSEF (Korea);
CONICET and UBACyT (Argentina);
FOM (The Netherlands);
STFC and the Royal Society (United Kingdom);
MSMT and GACR (Czech Republic);
CRC Program and NSERC (Canada);
BMBF and DFG (Germany);
SFI (Ireland);
The Swedish Research Council (Sweden);
and
CAS and CNSF (China).

\appendix
\begin{figure*}[h]
	\includegraphics[width=0.45\linewidth]{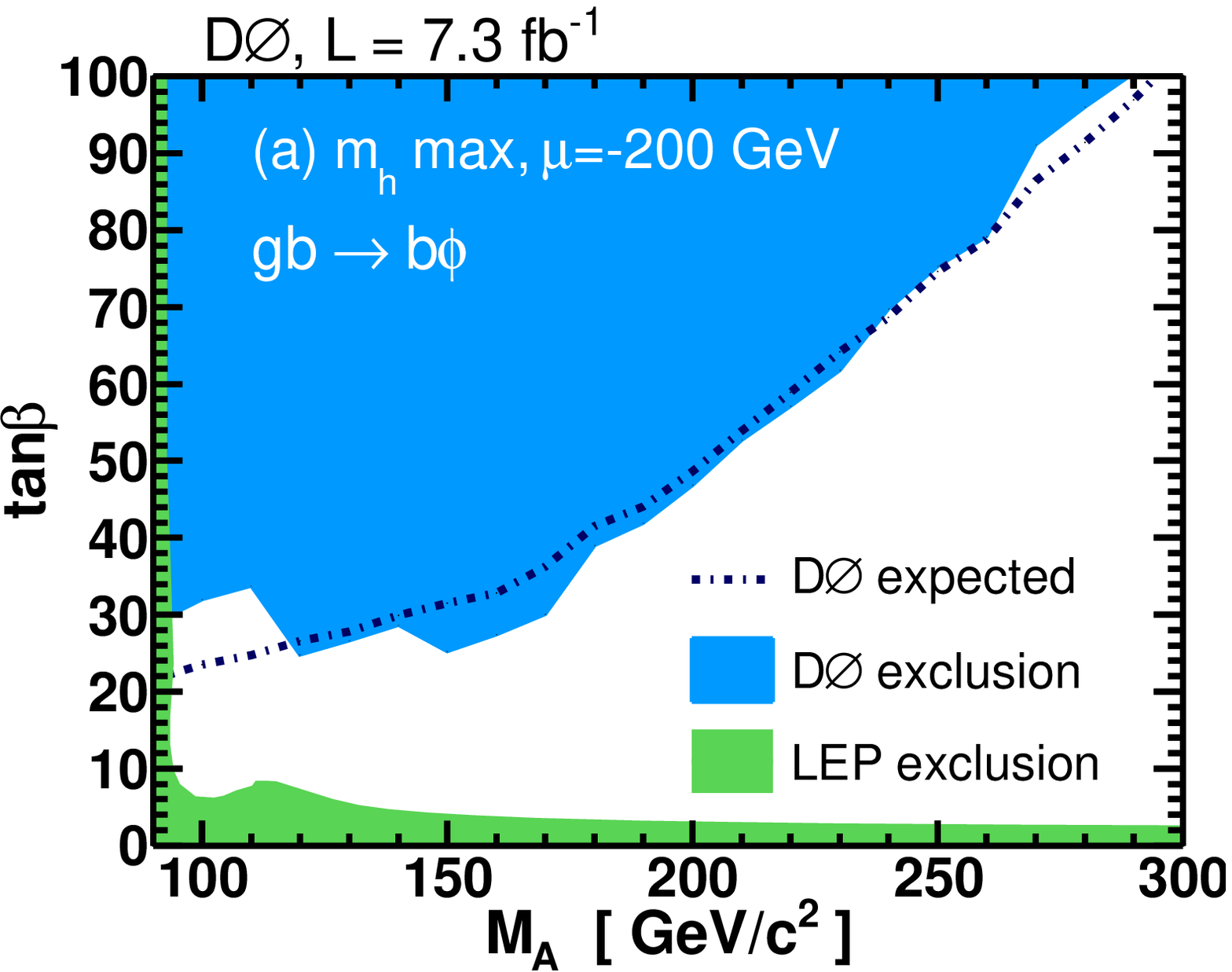}	
	\includegraphics[width=0.45\linewidth]{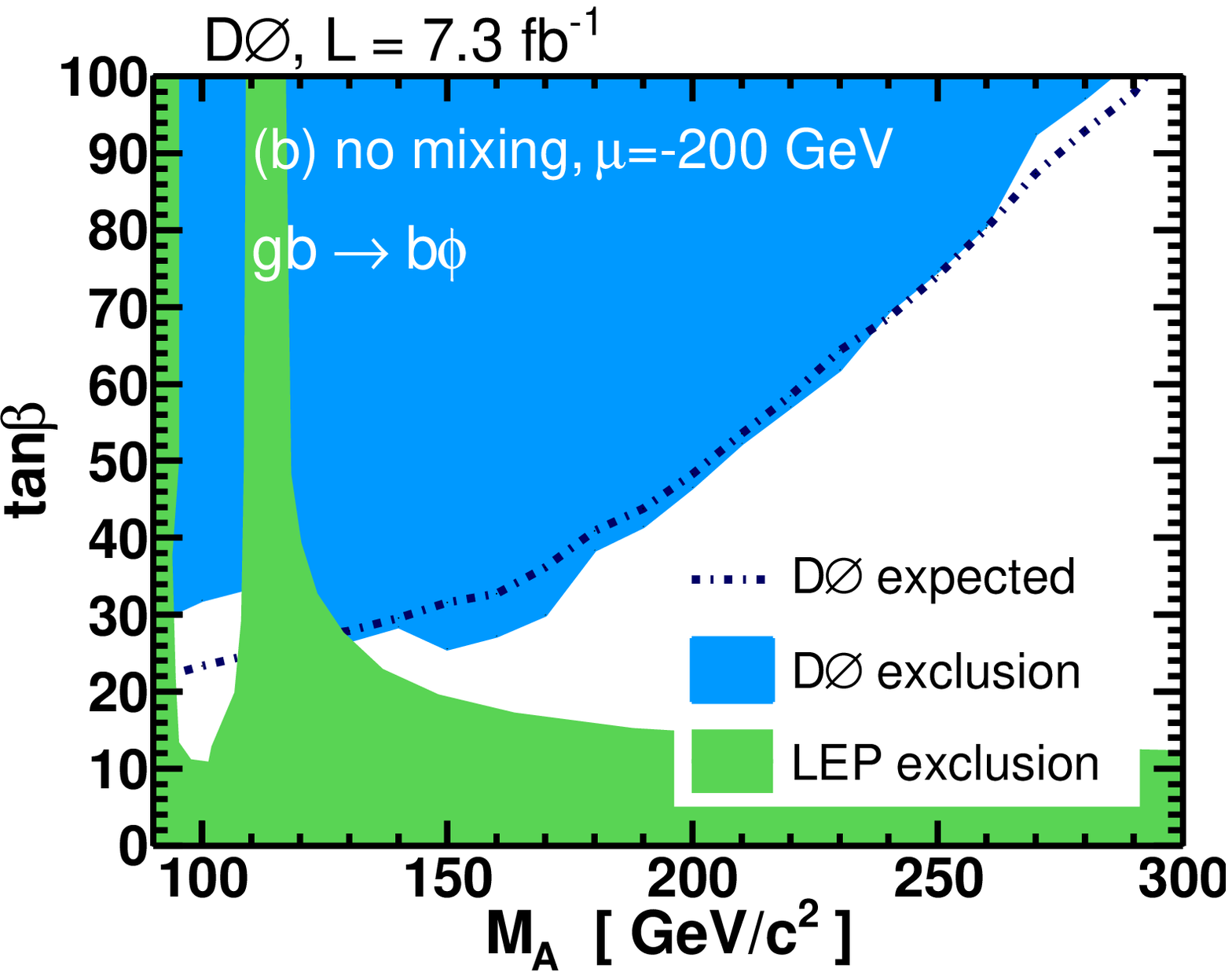}	
	
	\caption{ Limits on $\tan\beta$  vs $M_A$ for different benchmark scenarios: 
			(a) $m_h^{\text{max}}$ with $\mu=-200~\gev$,
			(b) no-mixing with $\mu=-200~\gev$.	
	}
	
	\label{fig2}
\end{figure*}

\begin{table*}[b]

	\caption{Expected and observed upper limits on $\tan\beta$ as a function of $M_A$ in four MSSM benchmark scenario.}
	\begin{ruledtabular}
		\begin{tabular}{c | rl | rl | rl | rl}

$M_{A}$& \multicolumn{2}{c}{$m_{h}$ max }& \multicolumn{2}{c}{$m_{h}$ max}  & \multicolumn{2}{c}{no mixing}& \multicolumn{2}{c}{no mixing}\\
($\gevcc$)               & \multicolumn{2}{c}{$\mu=-200~\gev$ }& \multicolumn{2}{c}{$\mu=+200~\gev$}  & \multicolumn{2}{c}{$\mu=-200~\gev$}& \multicolumn{2}{c}{$\mu=+200~\gev$}\\
             & Obs. & Exp.& Obs. & Exp.& Obs. & Exp.& Obs. & Exp.\\
\hline
$90$ & 28.9 & 21.6 & 28.8 & 21.3 & 29.0 & 21.6 & 28.8 & 21.2 \\
$100$ & 32.0 & 23.5 & 32.0 & 23.5 & 31.9 & 23.3 & 31.7 & 23.2 \\
$110$ & 33.8 & 24.7 & 33.6 & 24.4 & 33.6 & 24.9 & 33.5 & 24.4 \\
$120$ & 24.8 & 26.6 & 25.0 & 26.7 & 25.0 & 27.0 & 24.9 & 26.8 \\
$130$ & 26.7 & 27.8 & 26.3 & 27.5 & 26.7 & 27.9 & 26.6 & 27.6 \\
$140$ & 28.7 & 29.9 & 28.1 & 29.3 & 28.5 & 29.5 & 28.2 & 29.4 \\
$150$ & 25.3 & 31.5 & 25.3 & 31.4 & 25.6 & 31.6 & 25.3 & 31.3 \\
$160$ & 27.5 & 32.8 & 27.3 & 32.6 & 27.2 & 32.7 & 27.4 & 32.4 \\
$170$ & 30.1 & 36.3 & 29.8 & 36.0 & 30.0 & 36.2 & 29.5 & 35.8 \\
$180$ & 39.0 & 41.5 & 38.4 & 40.8 & 38.4 & 40.9 & 38.4 & 40.7 \\
$190$ & 41.9 & 44.1 & 41.6 & 43.4 & 41.6 & 43.8 & 41.5 & 43.6 \\
$200$ & 46.9 & 48.6 & 45.8 & 47.1 & 46.6 & 48.2 & 45.5 & 47.4 \\
$210$ & 52.8 & 53.9 & 51.8 & 53.3 & 52.3 & 53.5 & 51.9 & 53.1 \\
$220$ & 57.2 & 58.9 & 56.7 & 58.0 & 57.0 & 58.5 & 56.5 & 57.8 \\
$230$ & 61.8 & 64.2 & 60.9 & 62.9 & 61.9 & 64.4 & 61.4 & 63.7 \\
$240$ & 69.8 & 68.7 & 68.6 & 67.7 & 69.5 & 68.6 & 68.7 & 67.8 \\
$250$ & 75.3 & 74.6 & 75.5 & 74.8 & 74.7 & 74.1 & 72.9 & 72.0 \\
$260$ & 79.2 & 78.8 & 81.0 & 80.4 & 80.7 & 80.3 & 79.3 & 78.8 \\
$270$ & 91.1 & 86.4 & 93.5 & 88.8 & 92.5 & 87.4 & 91.1 & 86.5 \\
$280$ & 96.1 & 91.4 & 99.2 & 94.1 & 97.2 & 92.9 & 96.8 & 92.0 \\
	\end{tabular}
	
	\end{ruledtabular}
\end{table*}


\begin{thebibliography}{99}
\bibitem{pseudorapidity}  
The pseudorapidity $\eta$ is defined relative to the center of the detector as $\eta=-\ln[\tan(\theta/2)]$ where $\theta$ is the polar angle with respect to the proton beam direction.
\bibitem{mcol}  
$M_{\text{col}}^2 \equiv |\vpttau+\vptmu+\vmet|/\beta_T\times\left[ 1 - \beta_T^2 -    \beta_z^2 \right]$,
with $\beta_z = \tanh\left[   ( \eta^{\taum} + \eta^{\tauh} ) /2 \right]$ and
$\beta_T=-\sin\left(\varphi^{\taum} - \varphi^{\tauh}\right) / $  $ [ \cosh \eta^{\taum}\, \sin(\varphi^{\tauh}-\varphi_T)  - \cosh\eta^{\tauh}\, \sin(\varphi^{\taum} -  \varphi_T)  ]$, 
 where $\varphi_T$ is the \met azimuthal angle.
 $M_{\text{col}}$ is defined as 
 $M_{\text{col}}= sign(M_{\text{col}}^2)\times\sqrt{|M_{\text{col}}^2|}$.
\bibitem{suppl} Additional tables and figures are provided below.
\end{thebibliography}
\end{document}